\documentclass[pre,showpacs,amsmath,amssymb,floatfix,twocolumn]{revtex4}

\newcommand{\be}{\begin{equation}}
\newcommand{\ee}{\end{equation}}
\newcommand{\ba}{\begin{eqnarray}}
\newcommand{\ea}{\end{eqnarray}}
\newcommand{\rd}{\mathrm{d}}            
\newcommand{\re}{\mathrm{e}}            
\newcommand{\ri}{\mathrm{i}}

\newcommand{\te}[1]{\mbox{\boldmath $ #1 $}}
\newcommand{\tes}[1]{\mbox{\boldmath ${\it #1 }$}}
\newcommand{\bx}{\te{x}}
\newcommand{\bn}{\te{n}}
\newcommand{\bs}{\te{s}}

\newcommand{\bbe}{\te{e}}
\newcommand{\bu}{\te{u}}

\newcommand{\br}{\te{r}}

\newcommand{\bF}{\te{F}}
\newcommand{\bT}{\te{T}}
\newcommand{\bU}{\te{U}}
\newcommand{\bI}{\te{I}}
\newcommand{\bE}{\te{E}}
\newcommand{\bR}{\te{R}}

\newcommand\eg{\textit{e.g.}\ }
\newcommand\ie{\textit{i.e.}\ }

\usepackage{graphicx}
\usepackage{dcolumn}

\begin{document}

\title{Enhanced low-Reynolds-number propulsion in heterogeneous viscous environments}
\author {A.~M. Leshansky$^{1}$}
\email{lisha@technion.ac.il}

\affiliation{$^1$Department of Chemical Engineering, Technion, Haifa, 32000, Israel}

\date{\today}

\begin{abstract}
It has been known for some time that some microorganisms can swim faster in high-viscosity gel-forming polymer solutions. These gel-like media come to mimic highly viscous heterogeneous environment that these microorganisms encounter \emph{in-vivo}. The qualitative explanation of this phenomena first offered by Berg and Turner [Nature (London) \textbf{278}, 349 (1979)], suggests that propulsion enhancement is a result of flagellum pushing on quasi-rigid loose polymer network formed in some polymer solutions. Inspired by these observations, inertia-less propulsion in a heterogeneous viscous medium composed of sparse array of stationary obstacles embedded into an incompressible Newtonian liquid is considered. It is demonstrated that for prescribed propulsion gaits, including propagating surface distortions and rotating helical filament, the propulsion speed is enhanced when compared to swimming in purely viscous solvent. It is also shown that the locomotion in heterogenous viscous media is characterized by improved hydrodynamic efficiency. The results of the rigorous numerical simulation of the rotating helical filament propelled through a random sparse array of stationary obstructions are in close agreement with predictions of the proposed resistive force theory based on effective media approximation.
\end{abstract}


\maketitle

\section{Introduction}

In the past years there has been an increasing interest in propulsion on small scales, both theoretically and experimentally. Interest to some natural modes of low-Reynolds-number locomotion has a long record in applied mathematics \cite{lighthill75,childress81} and underlying mechanisms of propulsion powered by flexible elastic filament \cite{taylor51,hancock53,beating}, rotating helical flagellum \cite{hancock53,rotating,AR07}, beating cilia \cite{GL92}, surface distortions of non-flagellated squirmers \cite{ESBM96,SS96} and some others are quite well understood. Artificial nature-inspired propellers, powered by either beating or rotating filaments \cite{exper} were recently fabricated and tested vs. the theoretical predictions; performance of such devices was also studied numerically via particle-based algorithms \cite{particle}.

Theoretical work on zero-Reynolds-number locomotion strategies (that are not necessarily biomimetic) for artificial micro-swimmers has attracted some attention quite recently and several modes of propulsion, such as three-link Purcell's swimmer \cite{3link_swim}, its ``symmetrized" version \cite{AR08} and generalized N-link swimmer \cite{dAK04}, three-sphere propeller \cite{3sph_swim}, swimmer propelled by arbitrary non-retractable cyclic shape strokes \cite{shape_stroke}, two-sphere ``pushmepullyou" \cite{pmpu}, surface treadmilling \cite{LKGA07} and surface tank-treading \cite{LK08} and others, were proposed and studied in details. A comprehensive review that provides the reader with state-of-the-art in low-Reynolds-number locomotion can be found in \cite{LP09}.

However, the aforementioned works address the propulsion through an unbounded Newtonian viscous liquid. Some bacterial cells are used to swim though complex and highly heterogeneous viscous environments rather than Newtonian viscous liquids, such as marine water. For instance, clinically important spirochetes navigate efficiently through dense extracellular matrix in host tissues and cross the blood-brain barrier \cite{spirochet,RL00}. Moreover, prospective design of artificial microrobots capable of propulsion through soft tissues, digestion tract, spinal canal, etc. relies on ability to efficiently navigate through complex heterogeneous environment. Some experimental observations of biological propulsion in gel-like polymer solutions cannot be explained in the framework of standard theories. The traditional theories for bacterial motion through purely viscous response predict that the swimming speed (at constant torque produced by flagellar motors) monotonically decreases with viscosity. Although the naive intuition suggests that, for instance, a rotating helical filament would move like a corkscrew when propelled through a very viscous liquid, this is actually never the case, no matter how large is the viscosity, as the zero-Reynolds-number swimming is purely geometric \cite{purcell77}. However, the swimming speed of Pseudomonas aeruginosa in polyvinylpyrollidone solutions increases with viscosity up to a certain point and thereafter decreases \cite{SD74}. Other flagellation types of externally flagellated bacteria (\eg Bacillus megaterium, Escherichia coli, Serratia marcescens, Sarcina ureae, Spirillum serpens and Thiospirillum jenense) also exhibit an increase in swimming speed in viscous solutions \cite{SD74}.  The disagreement with traditional theories is more drastic for spirochetes, lacking external flagella, in which a helical or plane wave of the cell body moves backward yielding rolling of the cell body and forward propulsion.  A remarkable feature of spirochete motility is that cells swim faster in a high viscosity gel-like media than they do in low-viscosity aqueous media \cite{KD75,spiro_swim}. The swimming speed of Leptospira interrogans monotonically increases with viscosity in medium supplemented with methylcellulose until the viscosity exceeds 300 cP \cite{KD75}. In low-viscosity, aqueous medium, T. denticola is observed to rotate without translating \cite{RLKNGS97}.

Berg and Turner \cite{BT79} investigated the effect of viscosity on propulsion of various microorganisms and suggested that the propulsion enhancement in \emph{in-vitro} experiments can be caused by loose and quasi-rigid networks formed by entangled linear polymer molecules such as  methylcellulose. They argued that the enhanced propulsion is a result of flagellum pushing on the polymer network and, therefore, the propulsion resembles the motion of a screw boring through wood. Merely increasing the viscosity of the medium using a non-gel-forming sucrose polymer, such as Ficoll, did not enhanced motility, which indicated that propulsion enhancement may be dependent on viscoelastic rheology of the medium. However, recent findings of \cite{elastic}, who examined the effect of vicsoelasticity on propulsion for various rheological models of the liquid, showed that addition of elastic response to material constitutive equation does not yield the enhancement of propulsion. The propulsion through  viscoelatsic liquid is always hindered (for the prescribed swimming gait), both velocity- and efficiency-wise, when compared to locomotion through viscous (Newtonian) solvent of the same viscosity.

The first attempt to explain the propulsion enhancement theoretically, following \cite{BT79}, was made in \cite{MK02} for externally flagellated bacteria and later extended in \cite{NAGM06} to spirochetes. They suggested that the local viscous resistance to the motion of the flagellum normal to its surface is considerably higher than that corresponding to the tangential motion (note that in purely viscous liquid the ratio of viscous resistances of a slender body is $\sim$2). Indeed, the normal motion of the helical threads yields pushing on polymer aggregates surrounding flagellum, while tangential motions yield no perturbation to the microstructure. Therefore, in gel-like media with a sufficient traction, the flagellum is propelled much like a corkscrew, without slippage. On the other hand, increase in the drag on the large bacterium head results in decay of the propulsion speed starting at some viscosity in agreement with previous experiments. In accord with these arguments is also the monotonic increase of the swimming speed of the organisms lacking a large passive head, such as spirochetes. When surrounded by a dense gel-like environment, the helical or flat waves of their body - depending on the spirochete species - get sufficient traction to propel the cell body through the medium like a corkscrew, i.e. without slippage. If traction is not sufficient, then the cell body slips against the external medium and the cell translates more slowly. Although the model proposed in \cite{MK02,NAGM06} shows limited qualitative agreement with experimental results, it involves two unknown phenomenological parameters -- ``apparent viscosities" of unclear physical origin.

The aim of the present study is to develop a more rigorous theoretical framework of propulsion through heterogeneous viscous media. In some sense, the working hypothesis behind this theory follows the original Berg and Turner's proposal \cite{BT79}, \ie the flagellum is pushing on the stationary matrix of obstacles, whereas the interaction is indirect, \ie mediated by a viscous solvent. The theory is based on solutions of averaged equations of viscous flow through random sparse array of obstructions such as fibers or spheres, mimicking heterogeneous gel-like polymer solutions where dense cores of the microgel particles (or chain-like aggregates in methylcellulose solutions \cite{gotlieb}) are surrounded by a viscous solvent containing dissolved polymer chains \cite{gel_struct}. We are interested to address the effect of the sparse network of stationary obstacles (\eg fibers or spherical cores) embedded into a viscous incompressible Newtonian solvent on propulsion of a force-(torque-)free swimmer. The properly averaged hydrodynamic fields (\ie pressure and velocity) in such media are governed by the \emph{effective Brinkman media} approximation \cite{brinkman47},
\be
-\nabla p+\mu \Delta \bu=\mu \te{\kappa} \cdot \bu\:, \label{eq:brinkman}
\ee
where $\bu$ is the average velocity field satisfying the continuity equation for the incompressible liquid, $\nabla\cdot\bu=0$, $p$ is the average pressure field and $\te{\kappa}$ is the tensorial damping coefficient. The essence of this mean field model is that, on average, the fluid in proximity to a stationary obstacle experiences a damping body force proportional to the local velocity accounting for the influence of the neighboring objects on the flow. By averaged quantities we mean ensemble averages over all possible arrangements of the surrounding obstructions about a ``test" obstacle. Far from the ``test" obstacle the velocity gradients are weak and the equation (\ref{eq:brinkman}) reduces to the differential form of Darcy's equation describing the flow through porous media and thus $\te{\kappa}$ is taken to be the Darcy resistance (or inverse of permeability). Spatially isotropic random matrix of obstacles can be described by a scalar resistance, $\kappa_{ij}=\alpha^2\:\delta_{ij}$ with $\delta_{ij}$ being the identity tensor. Thus, $\alpha^{-1}$ defines a new length scale related to the stationary obstacles (usually referred as  ``screening" or shielding" length), so that for $r<\alpha^{-1}$ the velocity field is Stokesian, while for $r>\alpha^{-1}$ the velocity satisfies Darcy's equation. In particular, the fundamental solution of the singularly forced Brinkman equation (\ie flow driven by a point force $f_i$ acting at $\bx_0$) are identical to these corresponding to (Fourier-transformed) equations of oscillatory Stokes flow where the frequency parameter $\alpha^2=-\ri \omega/\nu$ replaces the hydrodynamic resistance \cite{KK91}
\be
u_i(\bx)=\frac{1}{8\pi\mu}\: \mathcal{G}_{ij}(\bx,\bx_0)\:f_i\:,\quad \mathcal{G}_{ij}=A\:\frac{\delta_{ij}}{r}+B\:\frac{\widehat{x}_i\:\widehat{x}_j}{r^3}\;, \label{eq:pointforce_brink}
\ee
with
\ba
A&=&2\re^{-\rho}\:(1+\rho^{-1}+\rho^{-2})-2\rho^{-2}\:, \nonumber \\
B&=&-2\re^{-\rho}\:(1+3\rho^{-1}+3\rho^{-2})+6\rho^{-2}\:,\nonumber
\ea
where $\widehat{\bx}=\bx-\bx_0$, $r=|\widehat{\bx}|$ and $\rho=\alpha r$. The fundamental solution for the pressure field corresponding to the Stokes flow \cite{KK91} remains unchanged. As expected, for $\alpha\rightarrow 0$ Stokes flow solution is recovered as $A,\:B\rightarrow 1$. Thus, the flow through random matrix of obstructions due to a point force decays like $r^{-3}$ (due to exponential screening of hydrodynamic interaction via $\re^{-\alpha r}$ in the above ``shielded Stokeslet" solution) vs. $r^{-1}$ decay of the classical point force (Stokeslet) solution in viscous  incompressible liquid.

The dumping coefficient ${\te \kappa}$ in (\ref{eq:brinkman}) can be deduced theoretically for some representative obstacle arrangements by relating it to the total drag on the individual elements comprising the network \cite{SG68, Howells74}. The calculation is based on the fact that the mean resistance force per unit volume is equal to (minus) the number density of obstructions times the mean drag force $\bF_1$ exerted on a single obstruction in the uniform flow $\bU$ though a sparse matrix, $\nabla p=n \bF_1=\mu \alpha^2\bU$. For the simplest model system of sparse random matrix composed of stationary spherical obstructions of radii $a$, so that $\bF_1=6\pi \mu a \bU$, we arrive at $\alpha^2=9\phi/2a^2$, where $\phi$ is the volume fraction or concentration of the obstacles. Self-consistent calculations of the components of $\te{\kappa}$ corresponding to fibrous media for various spatial arrangement of long fibers, can be found in \cite{SG68}.

It should be noted that effective media approximation (\ref{eq:brinkman}) was originally derived in \cite{brinkman47} from heuristic arguments. However, it was shown later to provide the leading correction to the Stokes drag on a test sphere for random distribution of stationary spherical obstacles in a rigorous theory where the localized resistance due to neighboring obstacles was taken into account in the multi-particle expansion scheme for small concentration of obstacles \cite{Howells74}. The excellent agreement between the second-order (in concentration) multi-particle theory and the theory based on (\ref{eq:brinkman}) suggested that the latter may serve as an accurate approximation even in the case of moderate obstacle concentration. The numerical results based on Stokesian Dynamics suggested that (\ref{eq:brinkman}) accurately describes the flow through the random arrays of spheres with up to 30\% (by volume) fraction of obstructions \cite{DHSK95}.

In this work we consider several locomotion modes extensively studied in the past and relevant for propulsion of flagellated as well as non-flagellated organisms. First, we consider the G.~I. Taylor's undulating deformable sheet propagating small-amplitude traveling waves along its surface  \cite{taylor51} and propelled through heterogeneous viscous medium whereas hydrodynamics is governed by (\ref{eq:brinkman}). The study of propulsion of non-flagellated swimmers powered by small-amplitude surface distortions (\ie squirmers) is greatly simplified via the use of Lorentz reciprocal theorem. Finally, we propose a local resistive force theory of propulsion through the effective media and test the predictions of the theory using rigorous numerical simulation of rotating helical filament through a sparse array of stationary spherical obstructions embedded within viscous Newtonian solvent.

\section{Propulsion of an undulating planar sheet}
\subsection{Transverse distortions}

\begin{figure}[t]
\includegraphics[scale=0.65]{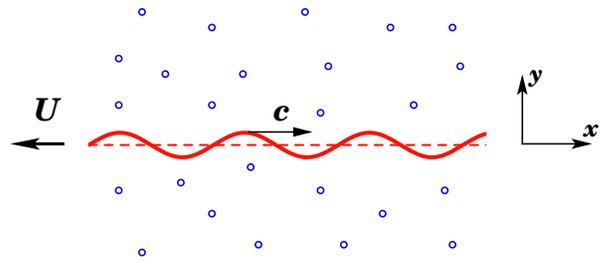}
\caption{(Color online) Schematic of the undulating sheet (red, solid) embedded in stationary random matrix of obstructions (blue dots).  The sheet is propagating a traveling transverse wave in the positive $x$ direction with the wave speed $c=\omega/k$ and propelled to the left with velocity $U$. \label{fig:schematic}}
\end{figure}

The 2D analog of the flagellum-powered swimming is the infinite sheet along which the traveling waves are propagating. G.~I. Taylor was the first to consider this model problem in the limit of long-wavelength (or, alternatively, small-amplitude) waves \cite{taylor51}. The same model was considered in \cite{elastic} for modeling propulsion in viscoelastic liquid. The extension of the asymptotic small-amplitude analysis towards the case when the flow around the sheet is governed by (\ref{eq:brinkman}) is rather straightforward. We first consider purely normal mode of deformation, so that in the reference frame fixed with sheet the position of the material points follow
\be
y_m=b \sin{(kx-\omega t)}\:,\qquad x_m=x\:,\label{eq:normal}
\ee
where $b$ is the amplitude of modulation, $k$ -- the wavenumber and $\omega$ is the frequency of the wave. The traveling wave is propagating in the positive x-direction with the wave speed $c=\omega/k$ (the schematic of the problem is shown in Fig.~\ref{fig:schematic}).
Since the problem is 2D we recast the equation of motions (\ref{eq:brinkman}) and the no-slip boundary conditions in terms of the streamfunction, $\psi$.
The streamfunction is defined by $u=\partial_y \psi\:,v=-\partial_x \psi$, where $\bu=\{u,v\}$ so that the incompressibility condition $\nabla\cdot \bu=0$ is trivially satisfied.
Assuming that the arrangement of the obstructions is random and isotropic (random arrangement of fibers, aligned in all directions or spherical obstacles in a random arrangement), the damping coefficient $\kappa_{ij}$ in (\ref{eq:brinkman}) is diagonal and equal to, $\alpha^2 \delta_{ij}$, where $\alpha$ has the dimensions of inverse length. Then, applying curl on both sides of (\ref{eq:brinkman}) yields
\be
\Delta \left(\Delta\: \psi-\alpha^2\:\psi\right)=0\:,\label{eq:brinkman1}
\ee
where $\Delta\equiv\nabla^2$.
The no-slip boundary conditions for the velocity components at the \emph{deformed} sheet read
\be
u=0\:,\qquad v=-b\omega\cos{(kx-\omega t)}\:.\label{eq:bc1}
\ee
Obviously, purely transverse surface distortions do not preserve surface area, as $\nabla_s\cdot\bu\ne0$, where $\nabla_s=(\te{I}-\bn\bn)\cdot\nabla$ is the surface gradient operator, and cannot describe deformation of flexible but incompressible sheet. In the latter case, the velocity distribution is given by a combination of transverse and tangential distortions. In the leading approximation the 2D incompressibility condition ($\partial u_s/\partial s+\kappa u_n=0$, $s$ is the arc length measured in the direction of tangent unit vector $\bs$) yields $\partial_x u+y''v=0$, that leads to \cite{taylor51}
\be
u=\frac{1}{4}b^2 k \omega \cos{(2kx-2\omega t)},\: v=-b\omega\cos{(kx-\omega t)}\:,\label{eq:bc1a}
\ee
We will discuss the propulsion by tangential distortions separately in the next section.

We expand the velocity at the deformed sheet in terms of its value at the \emph{undistorted} surface (at time $t=0$, the choice of time is arbitrary) and choose the following characteristic scales: $k^{-1}$ for length, $b\omega$ for velocity and $(b\omega k)^{-1}$ for time. In terms of the scaled variables we obtain the following problem for the nondimensional streamfunction
\ba
\Delta\left(\Delta-\alpha_*^2 \right)\psi=0\:, & \qquad y>0\:, & \label{eq:brinkman2}\\
\nabla \psi+\varepsilon \sin{x}\:\nabla \partial_y \psi=\cos{x}\: \bbe_x\:, & \qquad y=0\:,& \label{eq:bc2}
\ea
where $\alpha_*=\alpha/k$ is the scaled resistance and $\varepsilon=bk\ll1$ is the scaled amplitude of the wave. We now expand the solution in series of powers of $\varepsilon$, $\psi=\psi_0+\varepsilon \psi_1+\ldots$. Submitting this expansion into (\ref{eq:brinkman2}-\ref{eq:bc2}) and matching terms of the same powers $\varepsilon$ we can construct the asymptotic solution. The net displacement will be determined by the $x$-velocity component of the flow far from the sheet,
\be
u_\infty=\partial_y \psi\:,\quad y \rightarrow \infty\:.
\ee

In the leading order $\mathcal{O}(\varepsilon^0)$ (\ref{eq:brinkman2}-\ref{eq:bc2}) produces
\ba
\Delta\left(\Delta-\alpha_*^2 \right)\psi_0=0, &\qquad y>0, & \label{eq:brinkman3}\\
\nabla \psi_0=\cos{x}\: \bbe_x, &\qquad  y=0. &\label{eq:bc3}
\ea
The zeroth order solution is readily found as
\be
\psi_0=\left(A_0 \re^{-\sqrt{1+\alpha_*^2}\:y}+B_0 \re^{-y} \right)\,\sin{x}\:,\label{eq:psi0}
\ee
where the constants $A_0$ and $B_0$ are determined from (\ref{eq:bc3}),
\be
A_0=-\frac{1}{\sqrt{1+\alpha_*^2}-1}\:,\quad B_0=\frac{\sqrt{1+\alpha_*^2}}{\sqrt{1+\alpha_*^2}-1}\:.\label{eq:A0B0}
\ee
The propulsion velocity, as in the case of purely viscous liquid appears at the next order (because of symmetry $\varepsilon \rightarrow -\varepsilon$) as $\partial_y \psi_0\rightarrow 0$ as $y\rightarrow \infty$.

\begin{figure}[t]
\includegraphics[scale=0.85]{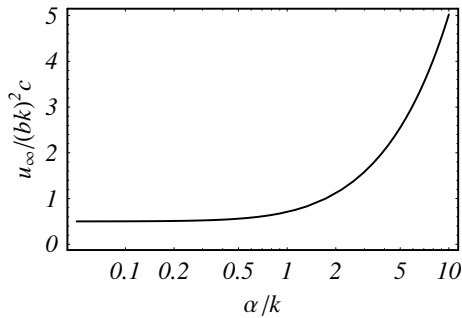}
\caption{Propulsion velocity $u_\infty/(bk)^2 c$ vs. the scaled mean resistance due to stationary obstructions, $\alpha/k$ (log-linear plot). The crossover from Stokesian regime to the enhanced regime dominated by the network hydrodynamic resistance occurs when screening length $\alpha^{-1}$ becomes comparable to the wavelength $k^{-1}$. \label{fig:Unormal}}
\end{figure}

At $\mathcal{O}(\varepsilon)$ we obtain the following problem
\ba
\Delta\left(\Delta-\alpha_*^2 \right)\psi_1=0\:, &\qquad y>0, &\label{eq:brinkman4}\\
\nabla \psi_1=-\sin{x} \nabla \partial_y \psi_0\:, &\qquad  y=0. &\label{eq:bc4}
\ea
Substitution of the zeroth order solution (\ref{eq:psi0}) into (\ref{eq:bc4}) at $y=0$ yields
\ba
\partial_y \psi_1&=&-\sin{x}\,\partial_{yy} \psi_0=\nonumber \\
&& -\left[(1+\alpha_*^2)A_0+B_0\right]\left(\frac{1}{2}-\frac{1}{2}\cos{2x}\right), \label{eq:bc4a} \\
\partial_x \psi_1&=&-\sin{x}\,\partial_{yx} \psi_0=0, \label{eq:bc4b}
\ea
as $\partial_y\psi_0=0$ at $y=0$. This suggests the form of the solution for $\psi_1$
\be
\psi_1=C_1 y+\left(A_1 \re^{-\sqrt{4+\alpha_*^2}\:y}+B_1 \re^{-2y} \right)\,\cos{2x}\:.\label{eq:psi1}
\ee
Substitution of (\ref{eq:psi1}) into (\ref{eq:bc4a}-\ref{eq:bc4b}) and using (\ref{eq:A0B0}) yields the unknown coefficients $A_1,\:B_1$ and $C_1$,
\[
A_1=-B_1=\frac{\sqrt{1+\alpha_*^2}\:\left(2+\sqrt{4+\alpha_*^2}\right)}{2\alpha_*^2},\; C_1=\frac{\sqrt{1+\alpha_*^2}}{2}\:
\]
The coefficient of the linear term, $C_1$, gives the velocity far from the undulating sheet,
\[
u_\infty=\left.\partial_y \psi_1\right|_{y\rightarrow\infty}=-\frac{1}{2}\left[(1+\alpha_*^2)A_0+B_0\right]=\frac{\sqrt{1+\alpha_*^2}}{2}.
\]
Thus, the sheet is propelled to the negative $x$ direction (propulsion velocity, $U$, is equal to minus $u_\infty$) as in the case of locomotion through viscous liquid without obstructions. Returning to the dimensional quantities, we obtain
\be
u_\infty=\frac{(bk)^2 c}{2}\: \sqrt{1+(\alpha/k)^2}\:, \label{eq:Unormal}
\ee
where $c=\omega/k$ stands for the speed of traveling wave. The classical Taylor's results is recovered at the limit $\alpha \rightarrow 0$. The propulsion velocity scaled with the wave speed is plotted in Fig.~\ref{fig:Unormal}. The crossover from Stokesian propulsion ($u_\infty=(bk)^2 c/2$) to that dominated by hydrodynamic resistance due to obstructions occurs at $\alpha_*\simeq 1$. In other words, propulsion powered by short wavelength surface distortions ($\alpha/k<1$) is not affected by the obstructions, while locomotion by long wavelength distortions (comparative to $\alpha^{-1}$) is enhanced, as the swimmer is experiencing the mean resistance due to stationary obstacles.

\subsection{Tangential distortions}
Let us consider now the effect of stationary obstacles on propulsion driven by purely tangential surface distortions (swimming gait of some non-flagellated microorganisms such as cyanobarcteria, \cite{ESBM96,SS96}). For this purpose we use the same model problem of the 2D sheet propagating traveling waves of compression/extension along its surface in the positive $x$ direction,
\be
y_m=y\:,\quad x_m=x+b \sin{(kx-\omega t)}\:,\label{eq:tang}
\ee
Similarly to the analysis of the transverse waves (Eqs. \ref{eq:brinkman2}-\ref{eq:bc2}), for purely tangential traveling waves we obtain the following problem in terms of the dimensionless stream function $\psi$ at the \emph{undeformed} surface:
\ba
\Delta\left(\Delta-\alpha_*^2 \right)\psi=0, &\qquad y>0, & \label{eq:brinkman5}\\
\nabla \psi+\varepsilon \sin{x}\:\nabla \partial_x \psi=-\cos{x}\: \bbe_y, &\qquad y=0. & \label{eq:bc5}
\ea

Substituting the asymptotic expansion $\psi=\psi_0+\varepsilon\psi_1+\ldots$ into (\ref{eq:brinkman5}-\ref{eq:bc5}) results at the
leading order $\mathcal{O}(\varepsilon^0)$ in the following boundary conditions
\be
\nabla \psi_0=-\cos{x}\: \bbe_y\:,\qquad  y=0\:,\label{eq:bc6}
\ee
while at any order $\psi_n$ should obey (\ref{eq:brinkman5}).

The zeroth order solution is
\be
\psi_0=\left(A_0 \re^{-\sqrt{1+\alpha_*^2}\:y}+B_0 \re^{-y} \right)\,\cos{x}\:,\label{eq:psi00}
\ee
where the constants are determined from by (\ref{eq:bc6}) and read
\[
A_0=-B_0=\frac{1}{\sqrt{1+\alpha_*^2}-1}\:.
\]
The net propulsion velocity is again restricted to the next order $\mathcal{O}(\varepsilon)$ as $\partial_y \psi_0\rightarrow 0$ as $y\rightarrow\infty$.

At $\mathcal{O}(\varepsilon)$ the boundary conditions for $\psi_1$ at $y=0$ read
\[
\nabla \psi_1=-\sin{x} \nabla \partial_x \psi_0\:,\qquad  y=0\:.
\]
Substituting the coefficients $A_0,\,B_0$ we find that at $y=0$
\ba
\partial_y\psi_1&=&-(\sqrt{1+\alpha_*^2}A_0+B_0)\sin^2{x}=-\left(\frac{1}{2}-\frac{1}{2}\cos{2x}\right), \nonumber \\
\partial_x\psi_1&=&(A_0+B_0) \cos{x}\sin{x}=0. \nonumber
\ea
These boundary conditions suggest the form of the solution for $\psi_1$ satisfying Eq.~\ref{eq:brinkman5}:
\[
\psi_1=-\frac{y}{2}+\left(A_1 \re^{-\sqrt{4+\alpha_*^2}\:y}+B_1 \re^{-2y} \right)\,\sin{2x}\:,
\]
and the remaining constants are readily found
\[
A_1=-B_1=-\frac{1}{2\sqrt{4+\alpha_*^2}-4}\:.
\]
The scaled fluid velocity at $y\rightarrow \infty$ is the same as for purely viscous liquid with viscosity $\mu$: $u_\infty=\partial_y \psi_1|_{(x,y \rightarrow\infty)}=-\varepsilon/2$, and being re-written in a dimensional form yields
\be
u_\infty=-\frac{c (bk)^2}{2}\:,
\ee
so the displacement is in the direction of wave propagation (positive $x$ direction). Therefore, the propulsion of the planer sheet powered by purely tangential small-amplitude surface distortions is not affected by presence of obstacles in the first approximation.

\subsection{Rate of viscous dissipation}
Since normal distortions are expected to facilitate propulsion, it is instructive to determine if such augmentation is costly in terms of power invested in swimming. The power expended by an arbitrary shaped organism and dissipated by viscosity is
\be
{\cal P}=-\int_S (\te{\sigma}\cdot\bn)\cdot \bu\: \rd S\:, \label{eq:dissp}
\ee
If decompose the surface velocity as $\bu=\bu'+\bU$ where $\bU$ is the propulsion  velocity of the swimmer and $\bu'$ is the velocity of the swimming stroke, and since $\te{\sigma}=\te{\sigma}'$ the integral in (\ref{eq:dissp}) remains unchanged when written in terms of $\bu'$ as the swimmer is force-free, $\int_S (\te{\sigma}\cdot\bn)\:\rd S=0$, (we omit the prime hereafter for simplicity). Re-writing the dissipation integral as ${\cal P}=2\mu \int_V \bE\,\te{:}\,\bE \: \rd V$, where $\bE$ is the rate-of-strain tensor and expressing the product $\bE \te{:}\,\bE$ as $\sum \zeta_i \zeta_i+2 (\partial_i u_j) (\partial_j u_i)$, where $\te{\zeta}=\mbox{curl}\bu$ denotes vorticity, allows expressing ${\cal P}$ as \cite{SS96}
\be
{\cal P}=\mu\int_V\te{\zeta}^2 \rd V-2\mu \int_S\:n_i\,(u_j\: \partial_j u_i) \:\rd S\:. \label{eq:dissp2}
\ee
Here $V$ is the fluid volume surrounding the swimmer.

For purely tangential surface distortions the second term on r.h.s of (\ref{eq:dissp2}) can be re-written as $2\mu \int_S\:\bu^2 \kappa_s\:\rd S$, where $\kappa_s=-(\partial \te{s}/\partial s)\te{\cdot}\:\bn$ is the curvature of the surface along the direction of the flow \cite{SS96}, and it is zero for un undistorted planar sheet. For purely transverse surface distortions the surface integral on the r.h.s. of (\ref{eq:dissp2}) is equal, to the first approximation, to $2\mu \int_S U \,\partial_x v\: \rd S$, where $U \sim c \varepsilon^2$ and $\partial_x v=\mathcal{O}(c \varepsilon k)$ and, therefore, its contribution to the dissipation rate is restricted to $\mathcal{O}(\varepsilon^3 \mu c^2)$. Thus, for both swimming gaits (normal and tangential surface waves), the rate-of-work, to the leading approximation, is given by the volume integral in (\ref{eq:dissp2}) over squared vorticity,  $\te{\zeta}^2=(\Delta\psi)^2$,
\be
{\cal P}=\mu\int_V \:(\Delta\psi)^2 \rd V\:. \label{eq:dissp3}
\ee
Substituting $\psi_0$ corresponding to purely normal (\ref{eq:psi0}) and purely tangential (\ref{eq:psi00}) gaits, respectively, and integrating over volume we find the rate of viscous dissipation to the leading approximation. It appears that to the leading approximation (with an error of $\mathcal{O}(\varepsilon)$) the dissipation rate (per unit area of the sheet) is the same for both swimming gaits and equal to
\be
\mathcal{P}=\mu c^2 \varepsilon^2 k \:\frac{2 + \alpha_*^2 + 2\sqrt{1 + \alpha_*^2}}{4\:\sqrt{1 + \alpha_*^2}}\:, \label{eq:dissp4}
\ee
where $\varepsilon=bk$. The ratio of rate-of-work (per unit area of the sheet) in propulsion through matrix of obstructions and that in unbounded viscous liquid ($\mathcal{P}_\mathrm{S}=\mu\: c^2 \varepsilon^2 k$ \cite{taylor51}, subscript `S' stands for ``Stokesian") $\mathcal{P}/\mathcal{P}_s \ge 1$ and it is a monotonically increasing function of $\alpha_*$. Again, the effect of the embedded obstruction matrix on rate of viscous dissipation is evident for distortions with wavelengths ($k^{-1}$) comparable to the screening length $\alpha^{-1}$.

\begin{figure}[t]
\includegraphics[scale=0.85]{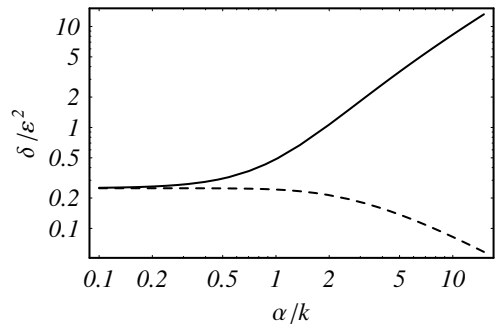}
\caption{Propulsion efficiency of the oscillating sheet, $\delta/\varepsilon^2$ vs. the dimensionless resistance, $\alpha/k$ (log-log plot): transverse surface waves (solid line); tangential (extension/compression) surface waves (dashed line). \label{fig:eff}}
\end{figure}

The routine definition of swimming efficiency based on the ratio between the work invested in dragging the immobile swimmer and that expanded in swimming (Lighthill's efficiency \cite{lighthill75}) is not applicable for 2D swimmers as the drag force is not defined due to Stokes paradox. This may be considered as a mere issue of normalization \cite{LKGA07,LK08} and a natural measure for the propulsion efficiency of an undulating sheet can be given by the ratio
\[
\delta=\frac{\mu U^2 k}{\mathcal{P}}\:.
\]
Using the expression (\ref{eq:Unormal}) for the propulsion velocity by the normal distortion and (\ref{eq:dissp4}) we arrive at
\be
\delta=\frac{\varepsilon^2\:(1 + \alpha_*^2)^{3/2}}{2 + \alpha_*^2 + 2\sqrt{1 + \alpha_*^2}}\:.\label{eq:eff}
\ee
The propulsion efficiency $\delta/\varepsilon^2$ in (\ref{eq:eff}) is depicted in Fig.~\ref{fig:eff} as a function of $\alpha_*$.
It is readily seen that the propulsion through matrix of obstructions is more efficient that swimming in the unbounded viscous liquid with $\delta_S=\varepsilon^2/4$.
We arrived at the improved propulsion efficiency for transverse distortions and retarded efficiency with longitudinal distortions. \emph{Therefore, propulsion through heterogenous viscous environment can be advantageous both speed-wise and efficiency-wise}. Although improved swimming speed is expected as a result of hydrodynamic interactions with obstacles for a prescribed swimming gait, the improved propulsion efficiency is rather surprising result. A similar occurrence of enhancement of dragging efficiency due to hydrodynamic interaction between a passive load and a micro-swimmer towing it was found in \cite{RL08}.

\section{Propulsion of non-flagellated squirmers by surface distortions}
\begin{figure}[t]
\includegraphics[scale=0.85]{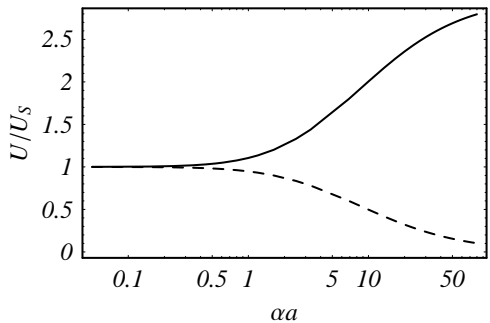}
\caption{Propulsion velocity of a spherical squirmer of radius $a$ moving through heterogeneous viscous medium as compared to Stokesian velocity of propulsion in unbounded viscous liquid, $U/U_\mathrm{S}$, vs. the dimensionless  resistance, $\alpha a$: purely normal surface distortions (solid line); purely tangential surface distortions (dashed line) \label{fig:squirmer}}
\end{figure}

We expect similar propulsion augmentation for non-flagellated swimmers (such as cyanobacteria \cite{ESBM96,SS96}) moving through the heterogeneous viscous media. In this case the analysis can be greatly simplified via the use of Lorentz reciprocity \cite{KK91}, that can be shown to hold for the effective media equations (\ref{eq:brinkman}) linear in $\bu$: for any two arbitrary solutions $(\bu,\te{\sigma})$ and $(\widehat{\bu},\widehat{\te{\sigma}})$ decaying far from the swimmer:
\be
\int_S (\widehat{\te{\sigma}}\cdot\bn)\cdot \bu\: \rd S=\int_S (\te{\sigma}\cdot\bn)\cdot \widehat{\bu}\: \rd S.\label{eq:recip}
\ee
Here $S$ is the instantaneous surface of the swimmer, $\te{\sigma}\cdot\bn$ is the local drag force the fluid exerts on $S$. The aim of this section is to show that the propulsion of cyanobacteria through viscous liquid with embedded network of obstructions can be enhanced when locomotion is powered by traveling \emph{normal} distortions of the outer surface. Following \cite{SS96} we will consider asymptotically tractable case of nearly spherical squirmer of radius $a$ propelled by small amplitude surface waves. Substituting $\bu=\bU+\bu'$, where $\bU$ is the propulsion velocity and $\bu'$ is the surface distortion in the frame of reference fixed with the squirmer's center-of-mass, into (\ref{eq:recip}) and setting the net force on the swimmer to zero, we arrive at
\be
\widehat{\bF}\cdot\bU\simeq-\int_S (\widehat{\te{\sigma}}\cdot\bn)\cdot \bu'\:\rd S\:,\label{eq:recip1}
\ee
where $\widehat{\te{\sigma}}$ is the stress field corresponding to the translation of the spherical squirmer when acted upon by an external force $\widehat{\bF}$. The equality in (\ref{eq:recip1}) is exact when the propulsion is governed by purely tangential distortions. For an auxiliary  problem of towing the immobile spherical squirmer through viscous liquid with embedded sparse network of obstructions, we use the well-known results for the drag force
\[
\widehat{\bF}=6\pi\mu a \left(1+\alpha a+(\alpha a)^2/9\right) \bU\,
\]
and
\[
\left.\widehat{\te{\sigma}}\cdot \bn\right|_S=-\frac{3\mu}{2a}\: \bU\cdot \left[(1+\alpha a)\bI+\frac{(\alpha a)^2}{3}\:\bn \bn \right]\:.
\]
for the local traction on the immobile squirmer's surface (\eg \cite{Howells74}). The resulting equation for the propulsion speed in terms of surface motions reads
\ba
\bU & \simeq &-\frac{1}{4\pi a^2 \left[1+\alpha a+(\alpha a)^2/9\right]}\times \nonumber \\
&& \:\int_S \left[(1+\alpha a) \bu'+\frac{(\alpha a)^2}{3} (\bu'\cdot\bn)\,\bn \right]\,\rd S\:. \label{eq:U}
\ea
The propulsion speed of a spherical squirmer in an unbounded viscous liquid is recovered in the limit $\alpha a \rightarrow 0$ in accord with \cite{SS96}, $\bU_\mathrm{S} \simeq -(4\pi a^2)^{-1} \int_S \bu' \rd S$ (the subscript ``S" stands for ``Stokesian"). Decomposing the arbitrary surface velocity into the normal and the tangential component, $\bu'=(\bI-\bn\bn)\cdot\bu'+(\bu'\cdot\bn)\,\bn$, allows Eq.~\ref{eq:U} to be re-written for either purely normal (superscript $n$) or purely tangential (superscript $s$) surface distortions as
\be
\bU^{n,s}=\mathcal{F}^{n,s}(\alpha a)\,\bU_\mathrm{S}\:,\label{eq:U1}
\ee
where $\mathcal{F}^{n,s}=\frac{\left\{1+\alpha a+(\alpha a)^2/3,\,\, 1+\alpha a\right\}}{1+\alpha a+(\alpha a)^2/9}$. Clearly, $1<\mathcal{F}^n<3$ and $0<\mathcal{F}^s<1$ for all values of $\alpha a>0$; the comparison of the propulsion velocity in a heterogeneous viscous medium and a purely viscous solvent for both kinds of surface motion (normal and tangential) is depicted in Fig.~\ref{fig:squirmer}.
Thus, when both swimming gaits (normal and tangential waves) of a squirmer propelled through heterogeneous viscous environment are compared to propulsion through unbounded viscous liquid, we find that normal surface distortions enhance the locomotion, while motion powered by purely tangential waves is hindered.

\section{Propulsion of a rotating helical filament}
\subsection{Modified resistive force theory for propulsion through heterogeneous viscous media}
\begin{figure}[t]
\includegraphics[scale=0.45]{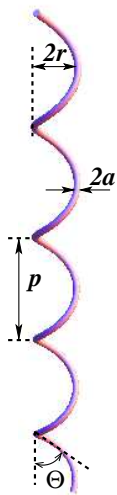}
\caption{(Color online) Schematics of a rigid helical filament: $2a$ is the diameter of the filament, $r$ is radius of the helix, $\theta$ is a pitch angle and a pitch is given by $p=2\pi r/\tan{\theta}$. \label{fig:helix_schematic}}
\end{figure}

In this section we develop a modification of the resistive force theory (RFT) of the propulsion through effective heterogeneous viscous media powered by rotating helical flagellum. We assume again that small inertia-less flow in such media is governed by (\ref{eq:brinkman}) and we aim to find the propulsion velocity and rate-of-work of the rotating rigid helix as a function of its geometry (\ie pitch angle $\theta$ and the radius $r$, see Fig.~\ref{fig:helix_schematic}) and angular velocity $\omega$.

The result for the propulsion speed of the force-free \footnote{The torque is not zero, however, it is a simplified model of a flagellum of externally flagellated microorganisms, the torque required for rotation is provided by the ATP-powered machinery within the cell body; the interaction with the cell body is neglected for simplicity} slender helix rotating in Newtonian incompressible viscous liquid reads (\eg \cite{AR07,WN08})
\be
\frac{U}{r \omega}= \frac{\sin{2 \theta}}{2\:(1+\sin^2{\theta})}\:,\label{eq:Uhelix}
\ee
so that the ratio of velocities of a helix to a corkscrew is independent of viscosity and by (\ref{eq:Uhelix}): $U/(r\omega\cot{\theta}) \le 1/2$. Thus, in the unbounded Newtonian viscous liquid the helix needs at least two turns to progress over a distance of its threads. The rate-of-viscous dissipation is given by
\be
\mathcal{P}=\frac{f_\perp L \: r^2 \omega^2}{1+\sin^2{\theta}}\:,\label{eq:dissp5}
\ee
where $f_{\perp}$ and $f_{||}$ the force densities corresponding to transverse and axial motion of a filament, respectively; $f_{\perp}\approx 2f_{||}$ for purely viscous liquid.

We are interested to extend these results and derive the propulsion velocity and power as a function of the hydrodynamic resistance of the random matrix of obstacles $\alpha$. Let us consider propulsion speed of a single rigid helix that rotates with some constant angular speed $\omega$. Taking $\bbe_3$ be the direction of the helical axis, the helix centerline is given by $\br(s,t)=\{r\cos{(k s+\omega t)},\,r\sin{(k s+\omega t)},\,b s + U t\}$, where $k=2\pi/\ell$ with $\ell$ being a length of a single helical turn so $kr=\sin{\theta}$; $b=\cos{\theta}$ and $U$ is yet undetermined propulsion velocity. The basic assumption of the local RFT, is that the local force per unit length (force density) exerted on the slender filament is given by
\be
\te{f}=f_{\perp}(\bu-(\bu\cdot\bs)\,\bs)+f_{||}(\bu\cdot\bs)\bs\:,\label{eq:rft}
\ee
where $\bs=\partial \br/\partial s$ is the local tangent, $\bu=\partial \br/\partial t$ is the local velocity and $f_{\perp}$ and $f_{||}$ the force densities corresponding to transverse and axial translation of a straight cylinder. For purely viscous liquid, $f_{\perp}=2f_{||}=4\pi\mu E$, with a small parameter $E=(\ln{2/\epsilon})^{-1}$ where $\epsilon=a/b\ll 1$ is an aspect ratio ($2b$ is the cylinder length and $2a$ is its diameter). The slenderness of the filament is controlled by the parameter $\kappa_f a\ll 1$, where $\kappa_f=|\partial^2 \br/\partial s^2|=\sin^2{\theta}/r$ is the local curvature of the filament centerline.

We substitute the local velocity as $\bu=\bu_\perp+U\bbe_3$ into (\ref{eq:rft}) and setting the total force in the $x_3$-direction to zero, we find that
\ba
U\int_0^L \left[f_\perp+(f_{||}-f_\perp (\bs\cdot\bbe_3)^2) \right] \rd s=\nonumber \\
\qquad -(f_{||}-f_\perp) \int_0^L (\bs\cdot\bu_\perp) (\bs\cdot\bbe_3) \rd s\:.\label{eq:rft1}
\ea
Substituting $\xi=f_\perp/f_{||}$ and the expressions for the local tangent $\bs$ and $\bu_\perp$ into (\ref{eq:rft1}) one can derive the propulsion velocity of the force-free helical swimmer rotating around its central axis with angular velocity $\omega$ as a function of $\xi$:
\be
\frac{U}{\omega r}=\frac{(\xi-1)\sin{2\theta}}{2\left[1+(\xi-1) \sin^2{\theta}\right]}\:.\label{eq:Uhelix2}
\ee
The RFT result (\ref{eq:Uhelix}) is readily recovered from (\ref{eq:Uhelix2}) for $f_{\perp}=2\,f_{||}$ and $\xi=2$. Similarly, the rate-of-dissipation $\mathcal{P}=-\int_0^L \te{f} \cdot \bu \rd s$, can be found
\be
\mathcal{P}=\frac{2 f_\perp L\: r^2 w^2}{1+\xi +(1-\xi ) \cos{2 \theta}}\:,\label{eq:power}
\ee
that reduces to the known RFT result (\ref{eq:dissp5}) for $\xi=2$. Also, the propulsion efficiency can be defined as a ratio of power required to drag the helical filament along the $x_3$-axis with velocity $U$ and the power expanded in force-free swimming, $\delta=\bF\cdot \bU/\mathcal{P}$. The force required to tow the helix is easily obtained by integrating the local force $\te{f}$ in (\ref{eq:rft}) for $\bu=U\bbe_3$:  $\bF=U\:(f_\perp (1-\cos^2{\theta})+f_{||}\:\cos^2{\theta})\:L \bbe_3$. Then, using $\bF$ and $\mathcal{P}$ yields an expression for  propulsion efficiency as a function of the pitch angle and the ratio $\xi$:
\be
\delta=\frac{(\xi-1)^2 \sin^2{2\theta}}{4\xi}\:.\label{eq:eff1}
\ee
For unbounded viscous liquid  $\xi=2$ and $\delta=\frac{1}{8}\sin^2{2\theta}$, yielding the optimal efficiency of 12.5\% at the pitch angle of $\theta=45^\circ$.

Now, let us solve the analogous problem for the force-free rotating helix propelled through a heterogeneous viscous medium. In this case (\ref{eq:rft1}) still holds, while the force densities $f_\perp$ and $f_{||}$ for a rigid cylinder of radius $a$ translating through medium with the effective resistance $\alpha$ are now functions of $\alpha a$ \cite{SG68},
\be
\frac{f_\perp}{4\pi\mu}=\frac{1}{4}\;(\alpha a)^2+\alpha a\;\frac{K_1(\alpha a)}{K_0(\alpha a)}\:,\qquad \frac{f_{||}}{4\pi\mu}=\frac{1}{2}\:\alpha a\;\frac{K_1(\alpha a)}{K_0(\alpha a)}\:,\label{eq:fbrink}
\ee
where $a$ is the radius of the filament and $K_p(x)$ are the modified Bessel functions of degree $p$. To determine the propulsion speed we only need to know the ratio of the resistance coefficients,
\be
\xi=\frac{f_\perp}{f_{||}}=2+\frac{\alpha a}{2}\;\frac{K_0(\alpha a)}{K_1(\alpha a)}\ge 2\:. \label{eq:xi}
\ee
Note that in the limit of vanishing matrix resistance $\alpha a\rightarrow 0$ the correct viscous limit, $\xi=2$, is recovered \footnote{However, the individual force densities $f_\perp$ and $f_{||}$ do not reduce to their corresponding Stokesian limiting values ($4\pi \mu E$ and $2\pi \mu E$, respectively) due to the fact that the limit $a\alpha \rightarrow 0$ is singular and requires special attention}. Substituting $\xi(\alpha a)$  into (\ref{eq:Uhelix2}) yields the expression for the propulsion velocity of slender helix rotating in viscous liquid with an embedded sparse matrix of obstacles. For vanishing resistance, $\xi \rightarrow 2$, and the propulsion velocity in (\ref{eq:Uhelix2}) tends to that in (\ref{eq:Uhelix}), as expected. In the opposite limit of large values of $\xi$ the motion resembles that of a corkscrew, $U=\omega r \cot{\theta}$, boring through solid without slip.
\begin{figure}[t]
\includegraphics[scale=0.85]{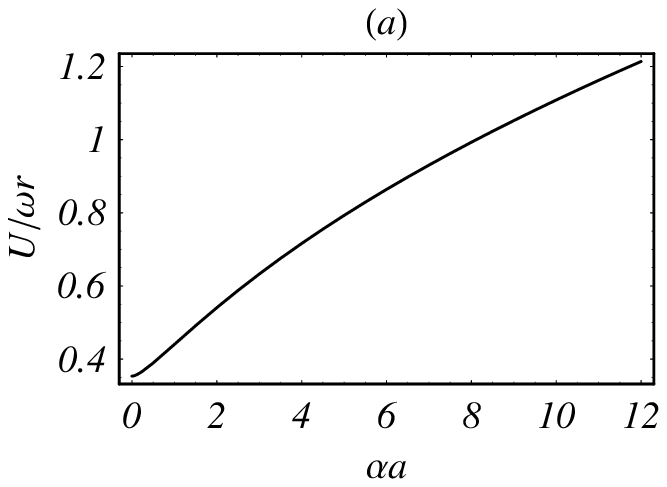}
\\
\includegraphics[scale=0.85]{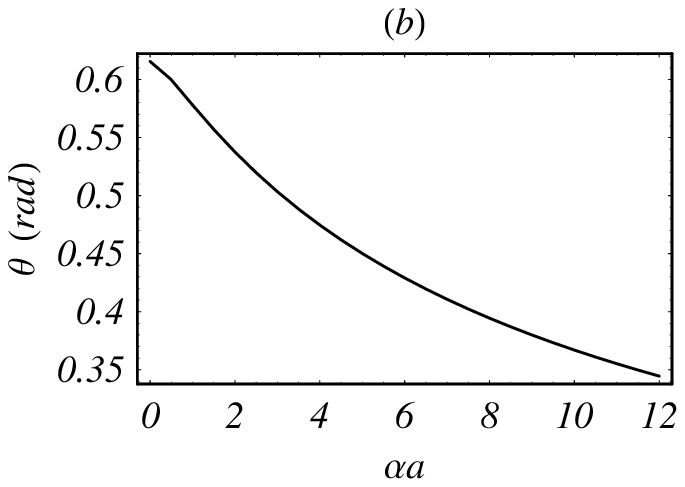}
\caption{Propulsion of a rotating helical filament through heterogeneous viscous media: (\emph{a})
optimal velocity, $U/\omega r$, vs. the scaled resistance, $\alpha a$; (\emph{b}) optimal pitch angle $\theta=\frac{1}{2}\arccos\left(\frac{\xi-1}{\xi+1}\right)$  vs. the scaled resistance $\alpha a$.
\label{fig:helix_vel}}
\end{figure}
Optimizing swimming speed for prescribed rotation velocity, we find that swimming speed is maximized at the pitch angle $\theta=\frac{1}{2}\arccos\left(\frac{\xi-1}{\xi+1}\right)$, vanishing resistance yields the pitch angle $\theta=\frac{1}{2}\arccos{\left(\frac{1}{3}\right)}\simeq 35.26^\circ$ that maximizes the propulsion speed in purely viscous liquid, giving $U/\omega r \simeq 0.354$ in agreement with previous theories (\eg see \cite{AR07}). The optimal propulsion velocity of the rotating helix moving through heterogeneous viscous environment is depicted in Fig.~\ref{fig:helix_vel}\emph{a} as a function of dimensionless hydrodynamic resistance $\alpha a$. The optimal pitch angle, however, decreases with the increase in $\alpha a$ (see Fig.~\ref{fig:helix_vel}\emph{b}).

\begin{figure}[t]
\includegraphics[scale=0.55]{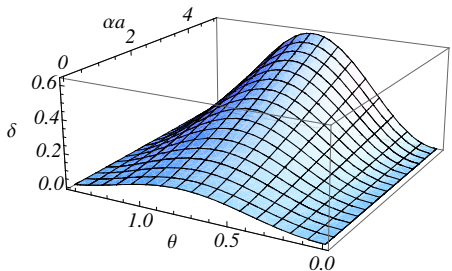}
\caption{(Color online) Propulsion efficiency $\delta$ (\ref{eq:eff1}) of a rotating helix in heterogeneous viscous media as a function of pitch angle $\theta$ (rad) and hydrodynamic resistance $\alpha a$. \label{fig:helix_delta}}
\end{figure}

Lastly, the efficiency of propulsion can be found by substituting $\xi$ in (\ref{eq:xi}) into (\ref{eq:eff1}). The resulting dependence on the scaled hydrodynamic resistance $\alpha a$ and the pitch angle $\theta$ is depicted in Fig.~\ref{fig:helix_delta}. It can be readily seen that, again, the propulsion is advantageous not only speed-wise, but also efficiency-wise as $\delta$ grows with $\alpha a$. Interestingly, the optimal pitch angle remains 45$^\circ$ for all values of $\alpha a$ since $\delta \propto \sin^2{2\theta}$. The enhanced efficiency predicted by the theory is not sufficient, however, for maintaining the swimming speed  for fixed torque applied to the filament. For a prescribed power expended in swimming, the swimming speed decays with $\alpha a$, which indicates that in addition to the proposed enhancement due to embedded matrix of obstacles, active control of motility by microorganisms may be involved.

\subsection{\label{sec:4b} Numerical simulations of propulsion through heterogeneous viscous medium}
\begin{figure}[t]
\includegraphics[scale=0.4]{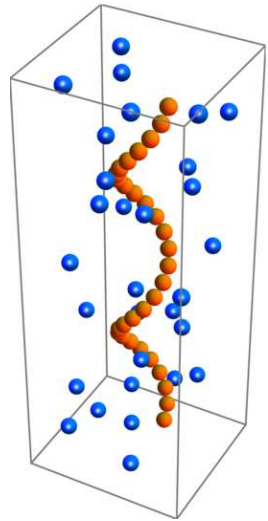}
\caption{(Color online) Illustration of a ``shish-kebab" filament propelled through heterogeneous viscous medium modeled as sparse matrix of stationary spheres immersed into Newtonian viscous liquid. \label{fig:helix_spheres}}
\end{figure}

To validate the theoretical predictions of the preceding section we implement the numerical scheme based on multipole expansion of the Lamb's  spherical harmonic solution of Stokes equations \cite{Filippov00}. The filament is constructed from nearly touching rigid spheres (``shish-kebab" model) and the obstruction matrix is modeled as random sparse array of stationary spheres (see Fig.~\ref{fig:helix_spheres}). All $N$ spheres (composing the filament and the matrix) have the same radius $a$. The no-slip condition at the surface of all spheres is enforced rigorously via the use of direct transformation between solid spherical harmonics centered at origins of different spheres. The method yields a system of $\mathcal{O}( N \mathcal{L}^2)$ linear equations for the expansion coefficients and the accuracy of calculations is controlled by the number of spherical harmonics (i.e. truncation level), $\mathcal{L}$,  retained in the series. The same approach was used in \cite{LK08,RL08} for modeling Purcell's toroidal swimmer.

The spheres composing the helical filament are partitioned along the backbone of the filament, $\br(s)=\left\{r\cos{(k s)},\,r\sin{(k s)},\,b s\right\}$, so that the distance between centers of neighboring spheres is set to $2.02 a$. Here $kr=\sin{\theta}$ and $b=\cos{\theta}$ with $r$ and $\theta$ being a radius and a pitch angle of the helix, respectively. The motion of the $i$th sphere composing a rotating helix can be decomposed into translation and rotation about its center as $\te{V}_i=\bU_i+\tes{\Omega}_i\times \br_i$ with $\bU_i=U\bbe_3+\omega\bbe_3 \times \bR_i$ and $\tes{\Omega}_i=\omega \bbe_3$; here $\bR_i$ is a position vector to the $i$th sphere center in the fixed laboratory frame and $\br_i$ is the radius vector with origin at the center of $i$th sphere, $\omega$ is the angular velocity of rotation and $U$ is the propulsion velocity along $x_3$-axis. The velocity of propulsion is determined by setting the net force exerted on rotating helix in the direction of propulsion to zero, $\sum_i {{F}_i}_3=0$, while translation and rotational velocities of the stationary obstacles are both set to zero. The rotation is powered by an external torque. The rate-of-work expended in propulsion of a rotating filament is found from
\ba
\mathcal{P}&=&\sum \limits_{i=1}^{N_p} (-\bU_i\cdot\bF_i-\tes{\Omega}_i\cdot \bT_i)= \nonumber \\
&& -\sum \limits_{i=1}^{N_p} \omega (-{R_i}_2\:{F_i}_1+{R_i}_1\:{F_i}_2+ {T_i}_3)\:,\label{eq:power1}
\ea
where $\bF_i=\int_{\partial S_i} \te{\sigma\cdot}\bn\:\rd S$ and $\bT_i=\int_{\partial S_i} \te{r}_i\times (\te{\sigma\cdot}\bn)\:\rd S$ are the hydrodynamic force and torque, respectively, exerted on $i$th sphere composing the filament.

\begin{figure}[t]
\includegraphics[scale=0.65]{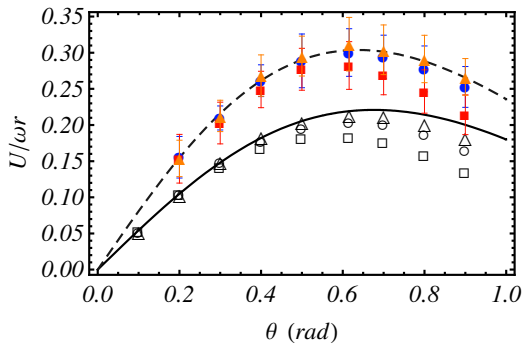}
\caption{(Color online) Propulsion velocity $U/\omega r$ of a rotating helical filament built of $30$ spheres  vs. the pitch angle $\theta$. The void symbols represent propulsion in unbounded viscous liquid: $r=2a$ ($\square$), $r=3a$ ($\bigcirc$) and $r=4a$ ($\bigtriangleup$); the full symbols stand for propulsion through viscous liquid with an embedded random matrix of 8\% (vol) stationary spherical obstacles: $r=2a$ ($\blacksquare$), $r=3a$ ($\bullet$) and $r=4a$ ($\blacktriangle$). The symbols stand for the mean value based on $20$ random configuration of obstacles; the error bars length is doubled mean standard deviation. The solid line is the theoretical result for propulsion in unbounded viscous liquid  (\ref{eq:Uhelix2}) (with $\xi=1.52$ as found from simulations for a ``shish-kebab" rod with $N_p=30$, see Fig.~\ref{fig:shishkebab}) and the dashed curve is the theoretical prediction (\ref{eq:Uhelix2}) for propulsion through sparse random matrix of obstructions with $\alpha a=0.6$ (corresponding to a random matrix of spherical obstacles with $\phi=0.08$) for $\xi(\alpha a)$ given in Eq.~\ref{eq:xi} and corrected for the ``shish-kebab" shape of the filament. \label{fig:helix_pitch}}
\end{figure}

\begin{figure}[t]
\includegraphics[scale=0.65]{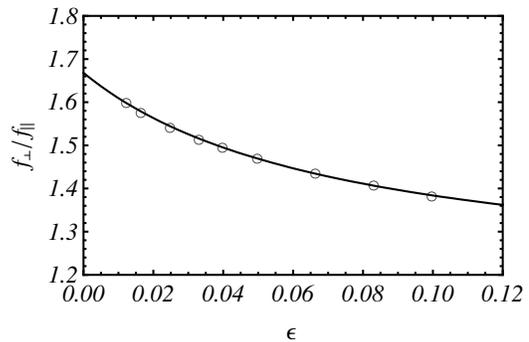}
\caption{Numerically calculated force ratio $f_{\perp}/f_{||}$ for a ``shish-kebab" straight rod of length $2b$ made of $N$ spheres of radii $a$ as a function of the aspect ratio $\epsilon=a/b$ (void symbols). The solid line stands for the best fit $(c_1+c_2 \epsilon)/(c_3+c_4\epsilon)$. \label{fig:shishkebab}}
\end{figure}

We initially test the scheme for the case of rotating helix moving through unbounded viscous liquid without obstacles. The results of the calculation are presented in Fig.~\ref{fig:helix_pitch} for a helix composed of $30$ spheres with radii $r=2a,\,3a$ and $4a$ upon varying the pitch angle $\theta$ (empty symbols).
The results of the calculation of the scaled propulsion velocity $U/\omega r$ vs. $\theta$ for small pitch angles (i.e. for $\kappa_f a=(a/r)\:\sin^2{\theta}\ll 1$, where $\kappa_f$ is a local curvature of the filament centerline) are in a good agreement with the RFT result (solid line) corrected  for a ``shish-kebab" shape of the helix. Actually, for such ``shish-kebab" filament variation of the shape occurs on a scale of the filament radius and not the length and therefore one should not expect the result of the local theory based on assumption of gentle variations of the shape, to hold in this case. However, this only changes the numerical coefficients in front of the resistance coefficients $f_\perp$ and $f_{||}$, leading to a slightly different ratio $f_\perp/f_{||}\approx 1.67$ at $\epsilon \rightarrow 0$ (instead of $\approx 2$ for slender particles with gradual variation of the geometry). The coefficient can be found numerically by computing the two force components, transverse and longitudinal, exerted on a ``shish-kebab" rod upon varying its length (\ie number of spheres composing the filament). The results of the calculation are presented in Fig.~\ref{fig:shishkebab}  together with the best fit of the form $(c_1+c_2 \epsilon)/(c_3+c_4\epsilon)$. The theoretical prediction of the helix velocity in Fig.~\ref{fig:helix_pitch} (solid curve) is based on (\ref{eq:Uhelix2}) with $\xi=f_\perp/f_{||}=1.52$ corresponding to a straight ``shish-kebab" filament composed of $N_o=30$ spheres (see Fig.~\ref{fig:shishkebab}).

It can be readily seen that the agreement between the numerical (void symbols) and theoretical result is very good for pitch angles $\theta<0.3$ rad, as the data corresponding to different helix radii ($r=2a$, $3a$ and $4a$) collapse on the theoretical curve (solid line in Fig.~\ref{fig:helix_pitch}). For higher pitch angle the assumption of slenderness ($\kappa_f a \ll 1$) is violated and deviation from the theory is evident due to hydrodynamic interaction of the helix threads. Increase in the radius of the helix (for the same pitch angle) yields better agreement with the RFT prediction for both: viscous liquid and heterogeneous viscous media. The position of the maximum is observed for the pitch angle $\theta\simeq 0.6$ (rad) in agreement with theoretical predictions regardless of the total length and radius of the helix. Increasing the number of spheres composing the helix results in slightly increased propulsion velocity, while the increase in accuracy level, $\mathcal{L}$, slightly diminishes the speed. Note that for slender helices the calculations are rather accurate even for just two harmonics ($\mathcal{L}=2$) retained in the series, as the hydrodynamic interaction between treads is negligible.

Next we consider propulsion through a sparse (8\% by vol) matrix of $N_o=30$ stationary spherical obstructions. Nonintersecting stationary spheres are arranged randomly in a rectangular box with its longer size oriented along the axis of the filament (see Fig.~\ref{fig:helix_spheres}). The size of the box is determined by the matrix density, i.e. volume fraction $\phi$ of the obstacles in the box and the slenderness of the filament. It should be realized, however, that the filament is propelled through a finite heterogeneous region surrounded by unbounded viscous liquid. However, the hydrodynamic disturbance from the rotating force-free filament in the heterogeneous effective media is expected to decay considerably faster due to ``shielded" interaction when compared to unbounded viscous liquid ($u=\emph{o}(1/r^3$), see Eq.~\ref{eq:pointforce_brink}), and thus the effect of the viscous domain outside the box is expected to be negligible. The propulsion velocity is determined from static rather than dynamic calculations in the same way as for filament rotating in purely viscous liquid. The value of the velocity is averaged over $20$ independent configurations and the results (full symbols) are shown in Fig.~\ref{fig:helix_pitch}. It can be readily seen that swimming through heterogeneous domain (see Fig.~\ref{fig:helix_pitch}) yields faster propulsion when compared to swimming in purely viscous liquid (for the prescribed swimming gait, \ie $\omega$). The agreement between the modified RFT (dashed curve) and the results of numerical simulations (full symbols) is very good. The theoretical prediction is based on (\ref{eq:Uhelix2}) with $\xi$ given by (\ref{eq:xi}) for $\alpha a=0.6$, corresponding to obstacle concentration of $\phi=0.08$ ($\alpha a=3\sqrt{\phi/2}$ for random array of spherical obstacles \cite{Howells74}), and multiplied by a factor of $0.83$ correcting for the ``shish-kebab" shape of the filament.

\begin{figure}[t]
\includegraphics[scale=0.65]{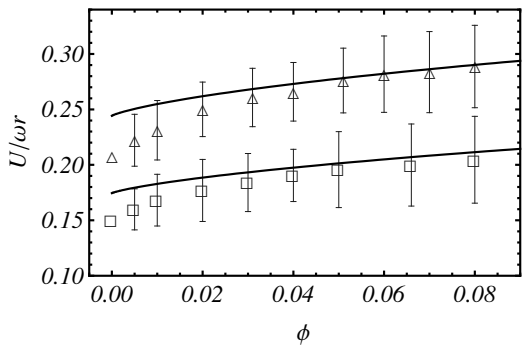}
\caption{Propulsion velocity of a rotating helical `shish-kebab" filament (as in Fig.\ref{fig:helix_spheres}) with $r=2a$, $\theta=0.3$ ($\square$) and $r=3a$, $\theta=0.5$ rad ($\vartriangle$) vs. concentration $\phi$ of stationary spherical obstructions, $\phi$. The solid curves are the modified RFT result (\ref{eq:Uhelix2}) with $\xi(\alpha a)$ given by (\ref{eq:xi}). Error bars show the doubled mean standard deviation of 20 random configurations. \label{fig:helix_conc}}
\end{figure}

The concentration dependence on the propulsion speed is depicted in Fig.~\ref{fig:helix_conc} for a helix composed of $N=30$ spheres with $r=2a$, $\theta=0.3$ and $r=3a$, $\theta=0.5$ for concentration of  spherical obstacles in a cell up to 8\% (vol). The monotonic increase in the propulsion speed $U/\omega r$ is evident and the agreement with the theoretical prediction (\ref{eq:Uhelix2}) is very good. The deviation from the theoretical prediction at small $\phi$ is probably due to the fact that derivation in \cite{SG68} leading to (\ref{eq:xi}) breaks down when the screening length and the filament length become comparable; this case requires a special consideration and will be addressed elsewhere.

\begin{figure}[t]
\includegraphics[scale=0.65]{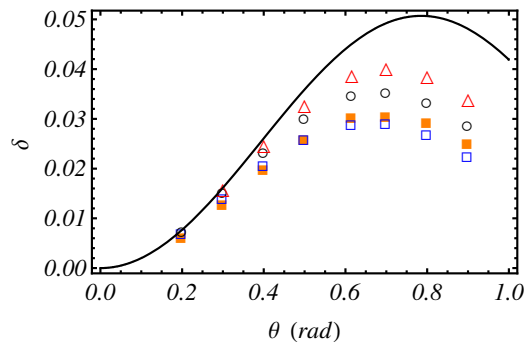}
\caption{(Color online) Efficiency of propulsion, $\delta$, of a rotating helical filament propelled through unbounded viscous liquid vs. the pitch angle $\theta$. Symbols stand for the results of the numerical calculation for a ``shish-kebab'' filament composed of $50$ spheres with: $r=6a$ ($\square$), $r=8a$ ($\circ$) and $r=10a$ ($\triangle$). Full squares ($\blacksquare$) are the results for a shorter helix with $N_p=30$ and $r=6a$. The continuous curve is a prediction of the RFT theory (\ref{eq:eff1}) with $\xi=1.56$ corresponding to the ``shish-kebab'' helix composed of $N_p=50$ particles. \label{fig:helix_eff1}}
\end{figure}

The power, $\mathcal{P}$, expended in propulsion  of the ``shish-kebab" filament is calculated via (\ref{eq:power1}). The corresponding hydrodynamic efficiency, $\delta$, is determined as $\delta=\mathcal{R}_{FU} U^2/\mathcal{P}$, where ${\cal R}_{FU}$ is the appropriate hydrodynamic resistance equal to the drag force on the (non-rotating) helix dragged along $x_3$-axis with $U=1$. The comparison of the numerical results for a helix composed of $50$ spheres and propelled through an unbounded viscous liquid, with the prediction of the RFT (\ref{eq:eff1}) corrected for the ``shish-kebab" shape (\ie $\xi=1.56$ corresponding to $N_p=50$) is depicted in Fig.~\ref{fig:helix_eff1} for several helix radii. We see that the theoretical prediction is less accurate than that for the propulsion speed (see Fig.~\ref{fig:helix_pitch}), and that the hydrodynamic interaction between treads at small $r$ diminishes the propulsion efficiency when compared to the theory. For $r=10\:a$ the agreement is good for pitch angles $<0.4$ rad. The position of the peak efficiency is slightly below the theoretical prediction of $\frac{\pi}{4}\approx 0.79$ rad and is around $0.7$ rad. Again, the increase in the radius of the helix (keeping helix radius fixed) results in a better agreement between the simulation results and the RFT prediction both for purely viscous liquid and heterogeneous media, as the slenderness parameter $\kappa_f a$ diminishes. The effect of the total length of the helix on propulsion efficiency is quite minor, a result for helix composed of $30$ spheres is very close to that of a longer helix with $N_p=50$ (see full squares vs. empty squares in Fig.~\ref{fig:helix_eff1}) for a wide range of the pitch angles.

\begin{figure}[t]
\includegraphics[scale=0.65]{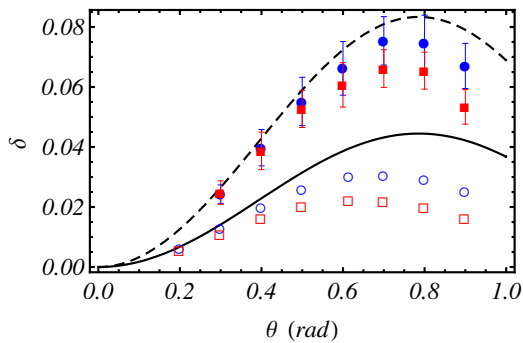}
\caption{(Color online) Efficiency of propulsion, $\delta$, of a rotating helical filament composed of $30$ spheres and propelled through viscous liquid with an embedded random matrix of 5\% (vol) stationary spherical obstacles, with vs. the pitch angle $\theta$. Full symbols stand for numerical results for propulsion heterogeneous viscous media with $5$\% (vol) of spherical obstacles embedded into viscous solvent: $r=4a$ ($\blacksquare$), $r=6a$ ($\bullet$); void symbols are the results of the numerical calculation for same filament, propelled through unbounded viscous liquid: $r=4a$ ($\square$), $r=6a$ ($\circ$).  Error bars depict the doubled mean standard deviation based on 20 random configurations of the obstruction matrix. The curves stand for predictions of the RFT theory using (\ref{eq:eff1}) for the propulsion efficiency in both cases: unbounded incompressible viscous liquid for $\xi=1.52$, corresponding to a helix composed of $N_p=30$ spheres (solid curve); heterogeneous viscous media using $\xi$ as in (\ref{eq:xi}) with $\alpha a=0.474$ corresponding to obstruction volume fraction $\phi=0.05$ (dashed curve).
\label{fig:helix_eff2}}
\end{figure}

Next we consider propulsion efficiency through sparse random array of stationary spherical obstructions immersed into incompressible viscous liquid. Here we use helices composed of $30$ spheres and obstruction matrix composed of $N_o=30$ spheres with volume fraction of obstacles of 5\% (vol). The results for the swimming efficiency are shown in Fig.~\ref{fig:helix_eff2} as full symbols for $r=4a$($\blacksquare$) and $r=6a$ ($\bullet$). In accord with the prediction of the modified RFT (the dashed line), the helix is a more efficient propeller in the presence of obstacles: the maximum efficiency increases more than two folds.  As for propulsion through purely viscous liquid (void symbols vs. solid line), the theoretical prediction based on effective media approximation overestimates the efficiency for large pitch angles, where hydrodynamic interaction of the treads is important, while the agreement is reasonably close for pitch angles $\theta<0.4$ rad. The RFT prediction was calculated from (\ref{eq:eff1}) using $\alpha a=0.474$ corresponding to obstacles' volume fraction of $\phi=0.05$, and $\xi$ multiplied by a constant factor of $0.83$ correcting for the ``shish-kebab" shape of the filament in the same way as done previously.

It should be stressed that no adjustable parameters are involved in comparison of the results of numerical simulations and the theory in Figs.~\ref{fig:helix_pitch}, \ref{fig:helix_conc}, \ref{fig:helix_eff1} and \ref{fig:helix_eff2} (besides the shape factor for the ``shish-kebab" filament that is estimated numerically).

\begin{figure}[t]
\includegraphics[scale=0.65]{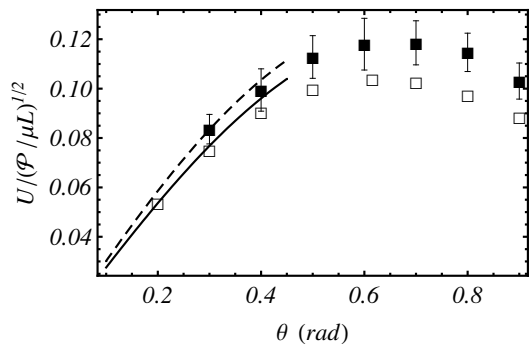}
\caption{Dimensionless propulsion velocity, $U/\sqrt{\mathcal{P}/\mu L}$, of a rotating ``shish-kebab" helix of radius $r=4a$ and $N=30$, vs. the pitch angle $\theta$. Full symbols ($\square$) stand for a helix propelled through a unbounded viscous liquid of viscosity $\mu$ and empty symbols, ($\blacksquare$), correspond to the helix propelled through heterogenous medium modeled as 5\% (vol) random array of stationary spherical obstructions of radius $a$ embedded into a viscous solvent of viscosity $\mu$. The solid and the dashed lines stand for the RFT result of propulsion in purely viscous solvent and viscous liquid with an embedded matrix of obstacles, respectively. \label{fig:fixpower}}
\end{figure}

Lastly, we calculate the propulsion velocity of a helix scaled with the rotation speed controlled by the power  invested in swimming (\ie torque applied to the filament), given by $\sqrt{\mathcal{P}/\mu L}$. The modified RFT theory (\ref{eq:Uhelix2},\ref{eq:power}) indicates that for a prescribed power the scaled propulsion speed, $U/\sqrt{\mathcal{P}/\mu L}$, should monotonically decay with the increase in $\alpha a$, while at $\alpha a \rightarrow 0$, this velocity grows unbounded as $f_\perp,\:f_{||}$ in (\ref{eq:fbrink}) vanishes (while their ratio $\xi$ remains finite). Note that $U/\sqrt{\mathcal{P}/\mu L}$ (in comparison to $U/\omega r$ and $\delta$ in (\ref{eq:Uhelix2}) and (\ref{eq:eff1}), respectively) is no longer a sole function of the ratio $\xi=f_\perp/f_{||}$, but also depends on the force density $f_\perp$ via $\mathcal{P}$. Obviously, as concentration of the obstacles tends to zero, it is expected that the force density $f_\perp$ should tend to a finite Stokesian value corresponding to an unbounded viscous liquid. This unphysical behavior near $\alpha a=0$ can be corrected by  constructing a proper slender body approximation for the Brinkman equation (\ref{eq:brinkman}) and will be addressed elsewhere. However, the growth of $U/\sqrt{\mathcal{P}/\mu L}$ as $\alpha a$ diminishes indicates the propulsion speed can be enhanced for a fixed power. For this matter, we compute the scaled propulsion speed of the rotating helix composed of $30$ particles and the embedded matrix composed of $N_o=30$ spheres with the volume fraction of obstacles of 5\%. The results of the computation are provided in Fig.~\ref{fig:fixpower} vs. the pitch angle, $\theta$. The numerical results (symbols) demonstrate the propulsion augmentation (up to $\sim 18$ \% of the Stokesian velocity) for rotating helix with fixed rate-of-work. The continuous lines refer to the modified RFT theory corrected for the shape of the filament as before, and the agreement with the numerical results is quite good for small pitch angles. Since the scaled propulsion $U/\sqrt{\mathcal{P}/\mu L}$ decays with $\alpha a$ and in the limit of vanishing resistance should yield a Stokesian result, it is expected that there should be the optimal resistance $\alpha a$ maximizing the propulsion speed of the helix.

\section{Concluding remarks}

We have demonstrated that transverse strokes/motions can lead to enhancement of propulsion via the heterogeneous viscous environment, modeled as sparse matrix of stationary obstacles embedded into incompressible viscous liquid, for different swimming gaits. The tangential motions,however, do not offer such enhancement. The present theory is in accord with the qualitative argument made in \cite{BT79} suggesting that ``...efficiency of flagellar propulsion is probably enhanced by the gel-like structure" of the medium. Combining the local theory and rigorous numerical calculations we demonstrated that the propulsion can be enhanced both speed- and efficiency-wise. For instance, the translation speed of the rotating helical filament is increased not only for a prescribed rotation velocity but also for a fixed torque applied to the filament. These findings can provide the physical grounds for explaining experiments showing enhanced propulsion of externally flagellated bacteria \cite{SD74,BT79} in viscous polymer gels.

We should next address the question of applicability of the effective media approximation to biologically relevant cases. Originally, the theory was developed to described the resistance to the flow through such media, i.e. stationary arrays of obstructions (fibers or spheres), while we consider a propulsion of a swimmer though such media. While the equations of motions (\ref{eq:brinkman}) remain invariant under Galilean transformation (up to a uniform pressure gradient), there is still an issue related to ``permeability" of such matrix to a finite-size object propelled through it \cite{Broday02}. The spacing between obstacles that compose the matrix places a constrain on the relevant dimension of the object moving through it without altering its structure. Obviously, for high volume fraction of the obstructions the effect of the matrix effect is most certainly local and not described by the effective mean resistance as in (\ref{eq:brinkman}). The question to what extent the Brinkman's approximation holds when the structure of the matrix is distorted in response to swimmer's movement, is yet to be answered. We can estimate the range of formal validity of the theory using some simple geometric arguments. For instance, for fibrous media (\eg chain polymer aggregates in gels) where the fiber axes are completely randomly oriented the value of $\alpha$ is determined in a self-consistent manner as function of mean fiber content \cite{SG68},
\[
\alpha_*^2=4 \phi\:\left(\frac{1}{3}\:\alpha_*^2+\frac{5}{6}\alpha_*\: \frac{K_1(\alpha_*)}{K_0(\alpha_*)}\right)\:,
\]
where $\alpha_*=\alpha b$, $\phi=\lambda \pi b^2$ is the fiber volume fraction with $\lambda$ being the mean fiber length per unit volume and $b$ stands for the fiber radius. Therefore, for sparse fiber matrix, $\phi=0.001$ ($0.1$\% vol), we find that $\alpha_*=0.0304$. Thus, for the typical value of $a/b=50$ (\eg the radius of the spirochete cell helix is $\sim 0.05$ $\mu$m, the typical radius of polysaccharide chains in the extracellular domain is $0.5-2$ nm) the corresponding value of $\alpha a=\alpha_* (a/b)=1.52$ and the optimal velocity of propulsion is $U/\omega r\simeq 0.49$ which is an improvement of  $\approx 40$\% comparative to the optimal speed of a helix propelled through viscous medium ($U/r\omega\simeq 0.35$). Assuming regular fiber spatial arrangements, one can estimate the average separation between fibers in the matrix $\Delta$ as
\[
\Delta/b \approx 2\left(\frac{\sqrt{3\pi}}{2}\:\phi^{-1/2}-1\right)\:.
\]
For $\phi=0.001$ this formula yields $\Delta/b\approx 100$, so that the mean spacing between fibers, $\Delta$, is sufficiently large to assume that the propelled helix is not distorting the matrix microstructure.

The situation is much more complicated when propulsion in gel-forming semi-dilute solutions of some polymers (such as methylcellulose) are considered. The typical ``mesh size" between polymer chain aggregates in these medium can be considerably higher than the above estimate based on uniform spatial arrangement, due to large concentration fluctuations typical for semi-dilute solutions of flexible and semi-flexible polymers \cite{DoiEdwards}. Therefore, the micro swimmer may navigate its way through polymer-lean region, taking advantage of the distributed hydrodynamic resistance due to clusters and/or aggregates in polymer-rich regions. The quantitative comparison of the present theory with the experimental results showing the significant increase of the propulsion velocity in gel-forming polymer solutions is difficult due to the complex microstructure of the gels, the local interaction of the propelled object with the polymer network, elastic response of the distorted network, etc. It should be further emphasized that the present work is not a quantitative description of propulsion in polymer gels, but it comes to qualitatively demonstrate the potential physical origin of propulsion enhancement due to screened hydrodynamic interaction with the embedded network.

The suggested numerical formalism of Sec.~\ref{sec:4b} can be readily extended to other cases of interest. If nearby obstacles are allowed to move in response to the propeller displacement when the certain stress threshold is exceeded, the rotating filament, for instance, could swim through arbitrary dense matrix of obstacles. Such approach can be applied for more realistic modeling of propulsion through viscoplastic materials exhibiting finite yield stress. A similar mechanism of locomotion enhancement is expected, as the swimmer is propelled through liquid-like region encapsulating the moving object, and it is ``pushing" against the unyielded solid bilk material. In the limit of small volume fractions of obstructions the effective media approximation (\ref{eq:brinkman}) should still hold, while in the other limiting case of freely suspended obstacles, the swimmer would experience hydrodynamic resistance controlled by the elevated effective viscosity, $\mu_\mathrm{eff}$, of the viscous suspension. Since zero-Reynolds-number propulsion through a viscous Newtonian liquid is purely geometric, propulsion velocity (for a prescribed swimming gait) is independent of the viscosity and no enhancement of locomotion is expected in this case. Adding elastic response of the distorted network of obstacles would allow more realistic modeling of propulsion in complex viscoelastic media such as viscous polymer gels.
\\

I would like to thank Oren Raz and Oded Kenneth for stimulating discussions on the subject. This work was supported by the Israel Science Foundation (Grant No. 923/07).


\begin{references}

\bibitem{lighthill75} J.~M. Lighthill, Mathematical Biofluiddynamics (SIAM, 1975).

\bibitem{childress81} S.~Childress, Mechanics of Swimming and Flying (Cambridge University Press, Cambridge, England, 1981).

\bibitem{taylor51} G.~I. Taylor, Proc. Roy. Soc. Lond. \textbf{A209}, 447 (1951).

\bibitem{hancock53} G.~J. Hancock, Proc. R. Soc. Lond. A 217, 96 (1953).

\bibitem{beating}  J. Gray and G.~J. Hancock, J. Exp. Biology \textbf{32}, 802 (1955); C.~H. Wiggins and R. E. Goldstein, Phys. Rev. Lett. 80, 3879 (1998); C. H. Wiggins et al., Biophys. J. \textbf{74}, 1043 (1998).

\bibitem{rotating} C. Brennen and H. Winet, Ann. Rev. Fluid Mech. 9, 339 (1977); E.~M. Purcell, Proc. Natl. Acad. Sci.USA 94, 11307 (1997).

\bibitem{AR07} O. Raz and J.E. Avron, New J. Phys. \textbf{9}, 437 (2007).

\bibitem{GL92} S. Gueron and N. Liron, Biophys. J. \textbf{63}, 1045 (1992).

\bibitem{ESBM96} K.~M. Ehlers, A.~D.~T. Samuel, H. C. Berg and R. Montgomery, Proc. Natl. Acad. Sci. USA \textbf{93}, 8340 (1996).

\bibitem{SS96} H.~A. Stone and A.~D.~T. Samuel, Phys. Rev. Lett. \textbf{77} 4102 (1996).


\bibitem{exper} R.~Dreyfus, et al., Nature \textbf{437}, 862 (2005); T.~S. Yu, E.~Lauga, and A.~E. {Hosoi}, Phys. Fluids \textbf{18}, 091701 (2006); B.~Behkam and M.~Sitti, J. Dyn. Syst. Meas. Control,  \textbf{128}, 36 (2006); K.~B. Yesin, K.~Vollmers, and B.~J. Nelson, Int.~J.~Robotics Res. \textbf{25}, 527 (2006); A. Ghosh and P. Fischer, Nano Lett., 2009, \textbf{9}, 2243 (2009); G. K\'osa, M. Shoham and M. Zaaroor,
IEEE Trans. Robot. \textbf{23}, 137 (2007).

\bibitem{particle} E. Gauger and H. Stark, Phys. Rev. E \textbf{74}, 021907 (2006); E.~E. Keaveny and M.~R. Maxey, J. Fluid Mech. \textbf{598}, 293 (2008).

\bibitem{3link_swim} L. E. Becker, S.~A. Koehler and H.~A. Stone, J. Fluid Mech. \textbf{490}, 15 (2003); D.~Tam and A.~E.~Hosoi, Phys. Rev. Lett. \textbf{98}, 068105 (2007); O.~Raz and J.~E. Avron, Phys. Rev. Lett. \textbf{100}, 029801 (2008).

\bibitem{AR08} J.~E. Avron and O. Raz, New J. Phys. \textbf{10}, 063016 (2008).

\bibitem{dAK04} G.~A. Araujo and J. Koiller , Qual. Theory Dyn. Sys. \textbf{4}, 139 (2004).

\bibitem{3sph_swim} A. Najafi and R. Golestanian, Phys. Rev. E \textbf{69}, 062901 (2004); R. Golestanian and A. Ajdari, Phys. Rev. Lett. \textbf{100}, 038101 (2008).

\bibitem{shape_stroke} A.~Shapere and F.~Wilczek, J. Fluid Mech. \textbf{198}, 557 (1989); J.~E.Avron, O. Gat and O. Kenneth, Phys. Rev. Lett. \textbf{93}, 186001 (2004).

\bibitem{pmpu} J.~E. Avron, O. Kenneth, and D.~H. Oaknin, New J. Phys \textbf{7}, 234 (2005).

\bibitem{LKGA07} A.~M. Leshansky et al., New J. Phys. \textbf{9}, 145 (2007).

\bibitem{LK08} A.~M Leshansky and O.~Kenneth, Phys. Fluids \textbf{20}, 063104 (2008).

\bibitem{LP09} E.~Lauga and T.~R. Powers, Rep. Prog. Phys. \textbf{72}, 096601 (2009).

\bibitem{spirochet} P. Coyle, and R. Dattwyler, Infect. Dis. Clin. North Am. \textbf{4}, 731 (1990); R.~Goldenberg, Am. J. Obstet. Gynecol. \textbf{189}, 861 (2003); N.~W. Charon and S.~F. Goldstein, Annu. Rev. Genet. \textbf{36}, 47 (2002).

\bibitem{RL00} J.~Radolf and S.~Lukehart, Pathogenic Treponema: Molecular and Cellular Biology (Caister Academic Press, Norfolk, England, 2006)

\bibitem{purcell77} E.~M. Purcell, Am. J. Phys. \textbf{45}, 3 (1977).

\bibitem{SD74} W.~R. Schneider and R.~N. Doetsch, J. Bacteriol. \textbf{117}, 696 (1974).

\bibitem{KD75} G.~E. Kaiser and R.~N. Doetsch, Nature, \textbf{255}, 656 (1975).

\bibitem{spiro_swim} E.~P. Greenberg and E. Canale-Parola, J. Bacteriol. \textbf{131}, 960 (1977); A. Klitorinos et al.,  Oral Microbiol. Immunol. \textbf{8}, 242 (1993); J.~D. Ruby and N.~W. Charon, FEMS Microbiol. Lett. \textbf{169}, 251 (1998);

\bibitem{RLKNGS97} J.~D. Ruby et al., J. Bacteriol. \textbf{179}, 1628 (1997).

\bibitem{BT79} H.~C. Berg and L.~Turner, Nature \textbf{278}, 349 (1979).

\bibitem{elastic} E.~Lauga,  Phys. Fluids \textbf{19}, 083104 (2007); H.~C. Fu, C. W. Wolgemuth and T.~R. Powers, Phys. Fluids \textbf{21}, 033102 (2009).

\bibitem{MK02} Y. Magariyama and S. Kudo, \texttt{Biophys. J.} \textbf{83}, 733 (2002)

\bibitem{NAGM06} S. Nakamura et al., Biophys. J. \textbf{90} 3019 (2006).

\bibitem{gotlieb} M. Gottlieb, personal communication.

\bibitem{gel_struct} G. P. Roberts and H. A. Barnes, Rheol. Acta \textbf{40}, 499 (2001); F. K. Oppong et al., Phys. Rrev. E \textbf{73}, 041405 (2006).

\bibitem{brinkman47} H.~C. Brinkman, Appl. Sci. Res. A \textbf{1}, 27 (1947).

\bibitem{KK91} S.~Kim  and S.~J.~Karrila,  \textit{Microhydrodynamics} (Butterworth--Heinemann, Boston, 1991).

\bibitem{SG68} L. Spielman and S.~L. Goren, Environ. Sci. Technol. \textbf{2}, 279 (1968).

\bibitem{Howells74} I. D. Howells, J.~Fluid Mech. \textbf{64}, 449 (1974).

\bibitem{DHSK95} T.~L. Dodd, et al., J. Fluid Mech. \textbf{293}, 147 (1995).

\bibitem{RL08} O. Raz and A. M. Leshansky, Phys. Rev. E \textbf{77}, 055305(R) (2008).

\bibitem{WN08} H.~Wada and R.~R. Netz, Phys.~Rev.~Lett \textbf{99}, 108102 (2007).

\bibitem{Filippov00} A.~V. Filippov, J.Colloid Interface Sci. \textbf{229}, 184 (2000).

\bibitem{Broday02} D. M. Broday, Bull. Math. Biol. {\bf 64}, 531 (2002).

\bibitem{DoiEdwards} M. Doi and S.~F. Edwards, The Theory of Polymer Dynamics (Oxford University Press, Oxford, 1988)


\end{references}
\end{document}